\documentclass[manuscript=article,layout=twocolumn,journal=acsaem]{achemso}

\usepackage{amsmath}
\usepackage{amssymb}
\usepackage{booktabs}
\usepackage[T1]{fontenc}
\usepackage{graphicx}    
\usepackage[utf8]{inputenc}
\usepackage[version=4]{mhchem}
\usepackage[sort&compress,numbers,super]{natbib}
\usepackage{setspace}
\usepackage{siunitx}
\usepackage{subcaption} 
\usepackage[table]{xcolor}

\usepackage[colorlinks=true,citecolor=blue,urlcolor=blue]{hyperref}
\usepackage[capitalise]{cleveref}

\author{Jeffrey S. Horner}
\affiliation{Thermal/Fluid Component Sciences Department, Sandia National Laboratories, Albuquerque, New Mexico, USA}
\author{Grace Whang}
\affiliation{Materials Science and Engineering Department, University of California, Los Angeles, Los Angeles, California, USA}
\author{David S. Ashby}
\affiliation{Quantum and Electronic Materials Department, Sandia National Laboratories, Livermore, California, USA}
\author{Igor V. Kolesnichenko}
\author{Timothy N. Lambert}
\affiliation{Photovoltaics and Materials Technology Department, Sandia National Laboratories, Albuquerque, New Mexico, USA}
\author{Bruce S. Dunn}
\affiliation{Materials Science and Engineering Department, University of California, Los Angeles, Los Angeles, California, USA}
\author{A. Alec Talin}
\affiliation{Quantum and Electronic Materials Department, Sandia National Laboratories, Livermore, California, USA}
\author{Scott A. Roberts}
\affiliation{Thermal/Fluid Component Sciences Department, Sandia National Laboratories, Albuquerque, New Mexico, USA} 
\email{sarober@sandia.gov}

\title{Electrochemical Modeling of GITT Measurements for Improved Solid-State Diffusion Coefficient Evaluation}

\setkeys{acs}{keywords = true}
\keywords{diffusion coefficient, lithium-ion battery, GITT, intercalation, \ce{FeS2}, electrochemical modeling}

\begin{document}


\begin{abstract}
Galvanostatic Intermittent Titration Technique (GITT) is widely used to evaluate solid-state diffusion coefficients in electrochemical systems. However, the existing analysis methods for GITT data require numerous assumptions, and the derived diffusion coefficients typically are not independently validated. To investigate the validity of the assumptions and derived diffusion coefficients, we employ a direct-pulse fitting method for interpreting GITT data that involves numerically fitting an electrochemical pulse and subsequent relaxation to a one-dimensional, single-particle, electrochemical model coupled with non-ideal transport to directly evaluate diffusion coefficients. Our non-ideal diffusion coefficients, which are extracted from GITT measurements of the intercalation regime of \ce{FeS2} and independently verified through discharge predictions, prove to be two orders of magnitude more accurate than ideal diffusion coefficients extracted using conventional methods. We further extend our model to a polydisperse set of particles to show the validity of a single-particle approach when the modeled radius is proportional to the total volume-to-surface-area ratio of the system.
\end{abstract}

\section{Introduction}
Critical scientific and social advancements, such as the widespread acceptance of electrical vehicles, are partly enabled by the precise calculation of battery material properties, such as the diffusion coefficient of lithium in the active cathode material\cite{Zhang2019,Rodrigues2017,Zeng2019}. However, reported diffusion coefficients for the same materials can vary drastically (often by orders of magnitude), depending on the particular experimental and/or theoretical model used to calculate them. For example, the ionic diffusion coefficient of lithium in \ce{Li2V2(PO4)3} has been reported with values ranging from \SI{e-15}{m^2/s} to \SI{e-12}{m^2/s}\cite{Ivanishchev2017}. These worrying discrepancies motivated us to look more closely at Galvanostatic Intermittent Titration Technique (GITT), which has been widely used to investigate solid-state diffusion in electrochemical systems. GITT subjects a cell to a series of constant current pulses that alternate between relatively short discharge/charge steps and relatively long rest steps\cite{Weppner1977}. Data gathered from the transient potential measurements can then be interpreted, using a model, to determine important physical parameters, including the reaction rate constant, Open Circuit Voltage (OCV), and diffusion coefficient\cite{Weppner1977,Verma2017}. However, while GITT provides useful data, interpreting that data through derived models has resulted in uncertain results. Recent studies have drawn attention to the fact that commonly applied analytical methods can result in predicted diffusion coefficients spanning orders of magnitude, and moreover, that GITT-derived diffusion coefficients seldom align with the diffusion coefficients obtained by other methods, including Potentiostatic Intermittent Titration Technique (PITT) or Electrochemical Impedance Spectroscopy (EIS)\cite{Nickol2020,Ivanishchev2017,Deng2020}. Nevertheless, insights gained from these experiments are essential for the accurate modeling of electrochemical systems -- a field that has been rapidly expanding throughout industry and academia in recent years\cite{Finegan2021,Lu2020,Mistry2021,Bielefeld2020,Ferraro2020}.

\begin{figure*}
    \includegraphics[width=\linewidth]{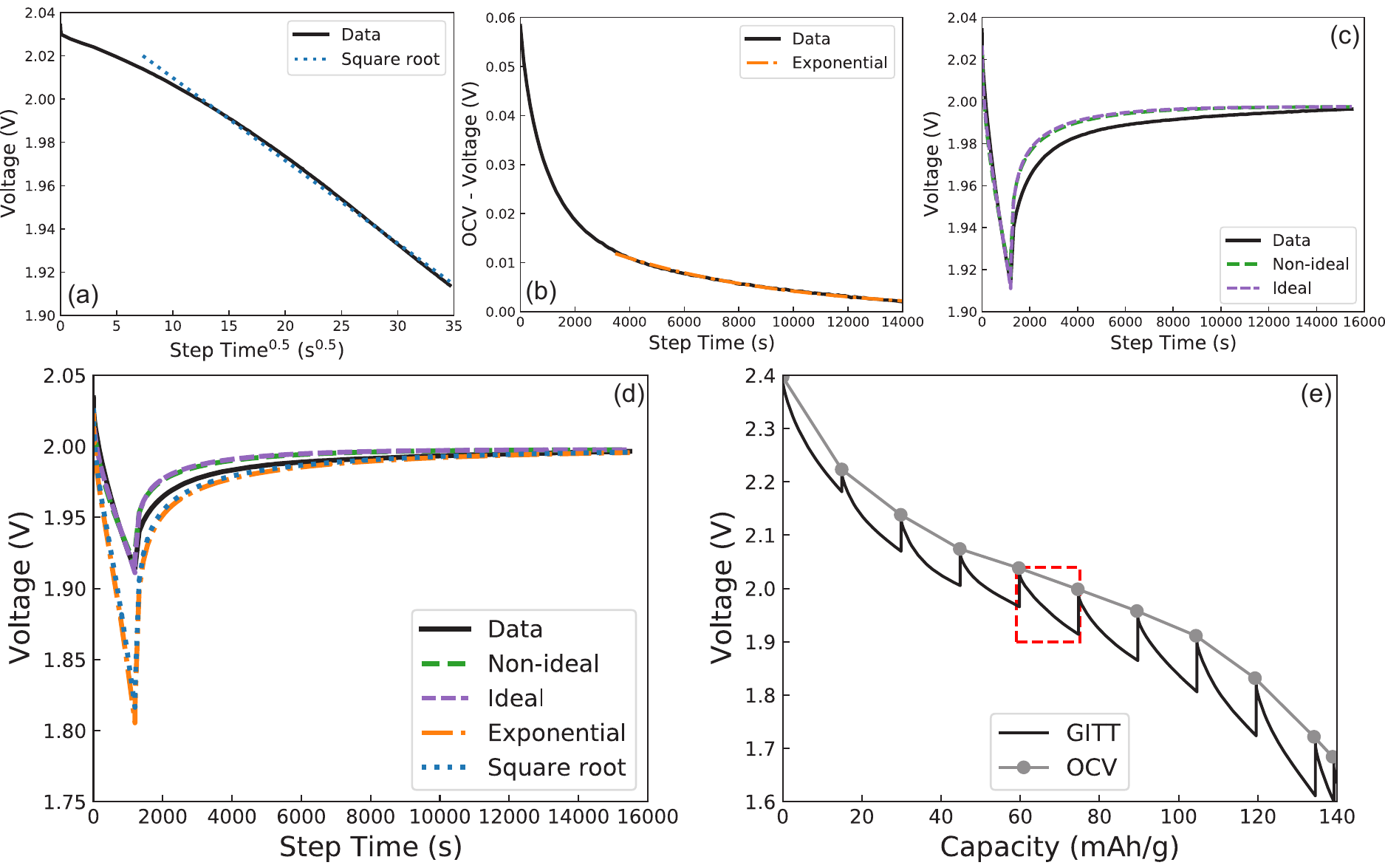}
    {
        \phantomsubcaption\label{fig:methods:sqrt}
        \phantomsubcaption\label{fig:methods:exp}
        \phantomsubcaption\label{fig:methods:direct}
        \phantomsubcaption\label{fig:methods:prediction}
        \phantomsubcaption\label{fig:methods:GITT}
    }
    \caption{\textbf{Methods for GITT interpretation.} (a) Square root fit method where the diffusion coefficient is determined through a linear fit of the charge/discharge voltage with respect to the square root of the step time. (b) Exponential fit method where an exponential is fit to the overpotential with respect to the step time. (c) Direct-pulse fitting method where the voltage is directly fit with a one-dimensional electrochemical model. (d) Resulting one-dimensional predictions from all methods for a single pulse. (e) GITT experimental data and OCV during lithiation at room temperature. The boxed step is the particular step considered in (a-d).}
    \label{fig:methods}
\end{figure*}

Diffusion coefficients are typically evaluated from GITT data through consideration of the charge or discharge steps. \citet{Weppner1977} originally derived a first-order approximation for a planar geometry, which suggests the potential $V$ response behaves linearly with respect to the square root of the step time (henceforth referred to as the ``square root fit'' and demonstrated in \cref{fig:methods:sqrt}). \citet{Delacourt2011} later proved this relation to also be accurate for spherical particles. With suitable knowledge of the active surface area $A_\mathrm{Tot}$ (a property that is typically difficult to measure and a potential source of error) and the OCV, one may compute the diffusion coefficient $D_\mathrm{Li}$ according to:
\begin{equation}
    D_\mathrm{Li} = \frac{4}{\pi}
    \left(\frac{i_\mathrm{Tot}}{F A_\mathrm{Tot}}\right)^2
    \left(\frac{d V_\mathrm{Eq}/d C_\mathrm{Li}}{d V/d \sqrt{t}}\right)^2,
    \label{eq:D_sqrt}
\end{equation}
where $F$ is the Faraday constant, $V_\mathrm{Eq}$ is the equilibrium potential, $C_\mathrm{Li}$ is the lithium concentration in the particles, and $t$ is the step time. This equation requires the assumption that the step time (i.e. the time during a pulse from when the current is switched on) is much less than the effective diffusion time $R_p^2/D_\mathrm{Li}$, where $R_p$ is the radius of the electrode particles. This equation also requires the contradictory assumption that the selected transient data must be at times large enough to not include the ohmic and kinetic overpotential, which creates an arbitrary selection criterion for data to be included in the analysis and results in different diffusion coefficient values that depend on the considered data range. 

Alternatively, it is also possible to extract diffusion coefficients from the rest portion of a GITT pulse, as has previously been done using a single exponential derived for planar conditions and a stretched exponential\cite{Amin2016,Jayawardana2018}. As shown in the Supporting Information, we develop a similar technique using a single exponential derived for spherical conditions at step times in excess of the effective diffusion time and when the potential is close to the equilibrium potential (henceforth referred to as the ``exponential fit'' and demonstrated in \cref{fig:methods:exp}). Although the square root fit and exponential fit make it possible to evaluate diffusion coefficients through linear regression, the credibility of these methods is questionable because they require researchers to assume criteria for the time points that cannot be known \textit{a priori}.

In addition, recent studies have estimated diffusion coefficients using direct-pulse fitting methods for a single-particle electrochemical model which enable consideration of an entire GITT pulse\cite{Shen2013,Hess2015}. For these models, the lithium flux into the particles $j_0$ is governed by the Butler-Volmer equation:
\begin{equation}
\begin{split}
    j_0 =  i_0 F & \left[ \exp\left(\frac{\alpha F \left(V_S - V_\mathrm{Eq}\right)}{R T}\right)\right. \\
    &-\left.\exp\left(-\frac{\alpha F \left(V_S - V_\mathrm{Eq}\right)}{R T}\right)\right],
    \label{eq:BV}
\end{split}
\end{equation}
where $\alpha$ is the charge transfer coefficient, $V_S$ is the solid-phase voltage, $R$ is the ideal gas constant, and $T$ is the temperature. The exchange current density $i_0$ is defined as:
\begin{equation}
    i_0 = k\left(C_\mathrm{Li^+}\right)^\alpha\left(C_\mathrm{max,Li}-C_\mathrm{Li}\right)^\alpha\left(C_\mathrm{Li}\right)^\alpha,
    \label{eq:ECD}
\end{equation}
where $k$ is the reaction rate constant, $C_\mathrm{Li^+}$ is the lithium-ion concentration in the electrolyte, $C_\mathrm{max,Li}$ is the maximum lithium concentration in the particles, and $C_\mathrm{Li}$ is the local lithium concentration in the particles. We use this approach for a single-particle model but also allow for the fact that lithium diffusion throughout the particle may either be ideal (i.e. Fick’s Law, where flux is driven by lithium concentration gradients) or non-ideal (i.e. concentrated solution theory, where flux is driven by potential gradients, a generalization of Fick's law). The direct-pulse fitting approach is advantageous as it requires fewer assumptions and uses the entire data provided in a pulse. A depiction of this fitting method is provided in \cref{fig:methods:direct} and a comparison of the resulting predictions for all methods using the single-particle electrochemical model are shown in \cref{fig:methods:prediction}. The predictions associated with the direct-pulse fitting methods result in a nearly two order of magnitude decrease in the discrepancy between the data and the model prediction. For reference, the entire GITT data used for this work is shown in \cref{fig:methods:GITT}, with the particular step considered in \cref{fig:methods:sqrt,fig:methods:exp,fig:methods:direct,fig:methods:prediction}, boxed.

In comparing the diffusion coefficients derived from common GITT analytical methods, we identify three common calculation assumptions that seem to be the source of potential errors in diffusion coefficient estimation: 1) existing GITT analytical approaches are only applicable to systems where the diffusive time scale is longer than the step time\cite{Weppner1977,Verma2017,Nickol2020}; 2) calculating the slope of the potential as a function of the square root of the step time applies at short times yet requires omission of ohmic and kinetic overpotentials, creating an arbitrary selection criteria\cite{Deiss2005}; and 3) all analytical methods for GITT were derived from ideal solution theory for diffusion (i.e. Fickian diffusion), even though intercalating electrochemical active materials are non-ideal, meaning the diffusion is driven by chemical potential gradients rather than concentration gradients\cite{Mendoza2016,Balluffi2005}. Consequently, we develop a new approach using the direct-pulse fitting method to calculating non-ideal diffusion coefficients from GITT measurements. We apply this method to the intercalation regime of \ce{FeS2} cathodes, a rechargeable battery material gaining interest due to its abundance\cite{Rickard2007}, low cost, non-toxicity, and high theoretical capacity (\SI{894}{mAh/g})\cite{Fong1989}. Furthermore, we demonstrate generality through applications to NCM523 (\ce{LiNi_{0.5}Co_{0.2}Mn_{0.3}O2}-based electrodes)\cite{Nickol2020}. Through our improved GITT analysis and a validation study, we demonstrate that non-ideal diffusion coefficients are fundamentally different than those used in ideal models and may be more than an order of magnitude less than ideal diffusion coefficients. We also explain the divergences and inaccuracies of the widely used diffusion coefficients obtained using existing GITT analytical methods. Furthermore, through an extension to polydisperse systems, we show that a single-particle approach is accurate if the effective radius is equal to three times the total volume-to-surface-area ratio of the system. Finally, we validate the newly obtained diffusion coefficients for our system against experimental discharge data and show that diffusion coefficients obtained through direct-pulse fitting of non-ideal solution theory result in a two order of magnitude reduction in the discrepancy with the data when compared to diffusion coefficients obtained through other means.

\section{Methods}

\subsection{Materials and Electrode Preparation}

\ce{FeS2} powder was purchased from Sigma Aldrich. Ball milling was performed in a Fritsch Pulverisette 7 Premium Line Planetary Micro Mill using a \SI{20}{mL} stainless steel Fritsch grinding bowl and \SI{3}{mm} stainless steel Fritsch media. In a typical run, 5-\SI{5.6}{g} of \ce{FeS2} were loaded into the grinding bowl along with an equivalent amount (by weight) of media and milled using one of the following two protocols: 1) 1,000 RPM for 3 hours without pausing, followed by a 5-10 minute rest, and again at 1,000 RPM for 3 hours without pausing or 2) 1,000 RPM for 6 hours without pausing. The contents of the grinding bowl were then sifted through nickel mesh to separate out the material from the media, before storing long-term in a glass vial in air.

\ce{FeS2} slurry electrodes were made by mixing ball-milled \ce{FeS2} powder, Super P (Alfa Aesar), and PVDF Binder (Kynar Flex 2801) in an 80:10:10 ratio by weight, respectively, in N-methyl-2-pyrrolidone (NMP, Sigma Aldrich), using a mortar and pestle to produce a homogeneous slurry. The slurry was subsequently doctor bladed onto a carbon-coated aluminum foil current collector (MTI Corp.) and dried in a vacuum oven overnight at \SI{120}{\celsius} prior to pumping into an Argon glove box (\textless1 ppm \ce{O2} and \textless1 ppm \ce{H2O}, VAC) for assembly into a coin cell. The thickness of the slurry is estimated from cross-section SEM (Supporting Information) to be approximately \SI{30}{\micro m}. Using the mass and density of the slurry added, we estimate the porosity to be 77\% for this uncalendared electrode, although we estimate high uncertainty in this value. Nevertheless, because we use a single-particle model with no ionic transport losses, the electrode thickness and porosity will not affect our results.

\subsection{Materials Characterization}

\begin{figure*}
    \centering
    \includegraphics[width=0.75\linewidth]{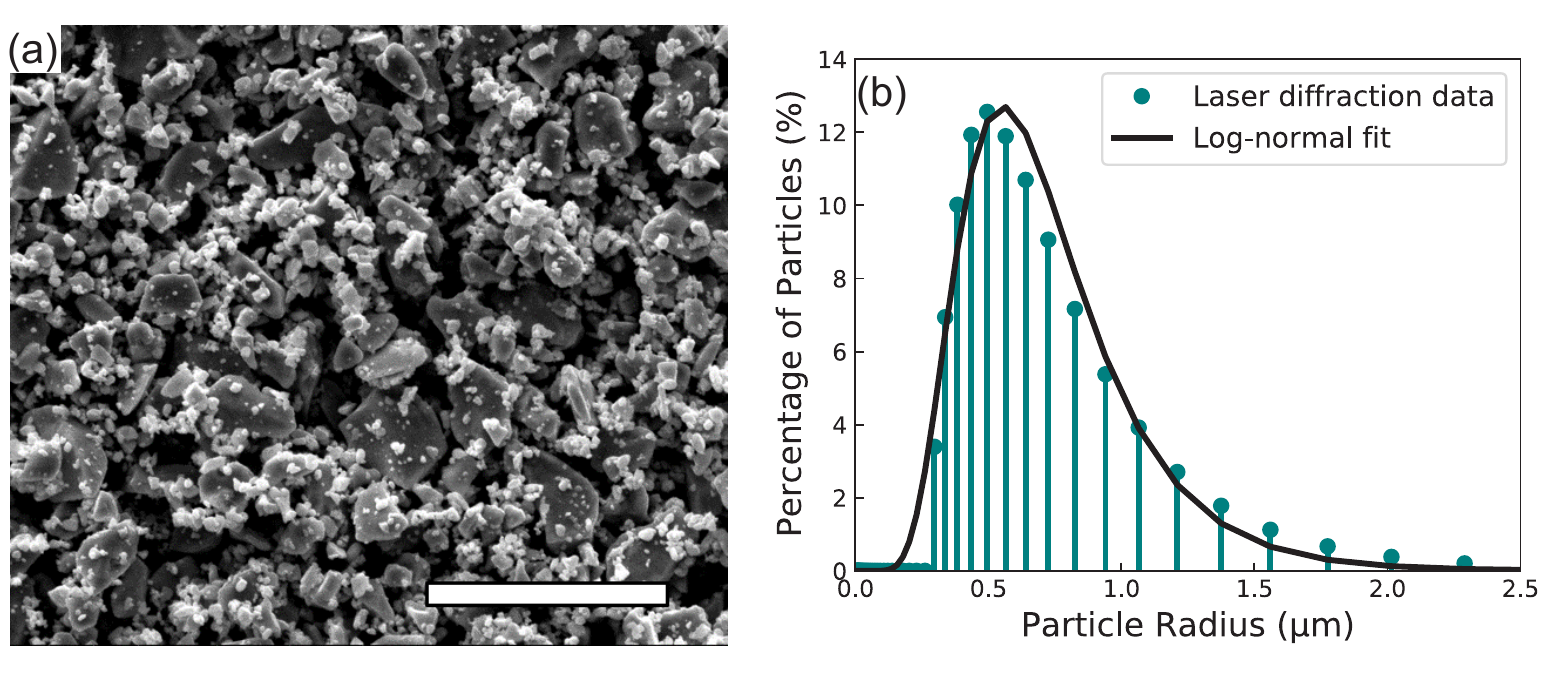}
    {
        \phantomsubcaption\label{fig:materials:SEM}
        \phantomsubcaption\label{fig:materials:PSD}
    }
    \caption{\textbf{Materials and electrochemical characterization of \ce{FeS2}.} (a) SEM of ball-milled \ce{FeS2} powders, demonstrating particle diameters under \SI{10}{\micro m}. Scale bar is \SI{10}{\micro m}. (b) Particle size distribution for ball-milled \ce{FeS2}, as determined through laser diffraction. A log-normal fit to the experimental data is provided.}
    \label{fig:materials}
\end{figure*}

Scanning electron microscope (SEM; Nova NanoSEM230; Thermo Fisher Scientific) images of the \ce{FeS2} powders were obtained using an accelerating voltage of \SI{10.0}{kV} and at a working distance of $\sim$\SI{5}{mm}. A representative image is shown in \cref{fig:materials:SEM}. Laser diffraction particle size analysis was performed using a Malvern Mastersizer 3000 instrument. The analysis was done using the dispersion unit accessory, with Vertrel\textsuperscript{TM} XF as the dispersing solvent. The refractive index was set at 3.08 and the absorption index at 0.01 before running the measurements, and then 1-5 mg of dry sample were added directly to the dispersion unit accessory already containing dispersing solvent. Solvent was flushed through the system between samples to prevent cross-contamination and faulty measurements. The resulting particle size distribution and log-normal fit are shown in \cref{fig:materials:PSD}.

\subsection{Electrochemical Characterization}

All electrochemical measurements in this study (GITT, discharge testing, and EIS) were performed on a BioLogic VMP3 potentiostat. \ce{FeS2} slurries were punched out into 3/8 inch diameter disks and assembled into a two-electrode 2032 coin cell with a lithium counter electrode, Celgard H1409 trilayer separator, and \SI{80}{\micro L} of \SI{1}{M} \ce{LiFSI} \ce{PYR14TFSI} electrolyte. The active mass loadings used in this study were $\sim$1-\SI{1.5}{mg/cm^2}.

\ce{FeS2} is a conversion cathode material that undergoes a four-electron redox process to produce \ce{Li2S} and \ce{Fe}. Prior to undergoing the conversion reaction, a significant portion of the total capacity results from lithium intercalation into the active material\cite{Fong1989,Evans2014,Zou2020,Yersak2012,Zhang2015}. To study the intercalation region, \ce{FeS2} coin cells were first lithiated down to \SI{1}{V} at a slow rate of C/50 to form \ce{Li2S} and \ce{Fe}, and then subjected to a partial delithiation up to \SI{2.4}{V} to avoid the upper conversion reaction\cite{Zou2020}. At \SI{2.4}{V}, the system is expected to be composed of \ce{Li_{1.2}FeS2}, the delithiated state of the intercalation reaction\cite{Fong1989,Son2013}. From this point on, the cycling window was limited to 1.6-\SI{2.4}{V} and GITT was performed at C/20 (based off a theoretical capacity of \SI{894}{mAh/g})\cite{Fong1989} for 20 minutes followed by a 4-hour rest period (\cref{fig:methods:GITT}). We provide the cycling performance of the cell in this limited voltage window in the supporting information. We performed the GITT experiments at both room temperature and \SI{60}{\celsius}, with the results for the \SI{60}{\celsius} experiments shown in the Supporting Information. For the intercalation region C-rate testing, \ce{FeS2} coin cells were subjected to the same lithiation and partial delithiation pre-cycling treatment and subsequently cycled within 1.6-\SI{2.4}{V} window.

\subsection{OCV Evaluation}

The OCV used throughout this work was evaluated from the GITT rest steps. As the measured voltage had not fully plateaued, an exponential was fit to the overpotential and extrapolated to determine the OCV for each step:
\begin{equation}
    \ln\left(OCV-V\right) = k_1-\frac{t}{\tau},
    \label{eq:OCV}
\end{equation}
where $k_1$ and $\tau$ are constants. Data at step times above 1 hour were only considered for the exponential fitting. The fit was performed by linearizing the exponential and evaluating the relaxation time and exponential prefactor through linear regression, whilst leaving the OCV (on which the overpotential depends) as a single nonlinear parameter to be fit. The nonlinear regression was performed using MATLAB’s built in function, \texttt{fminsearch}, with the sum of the squared potential residuals as the objective function to be minimized. If a non-monotonic behavior was observed in the rest step potential, the final measurement was assumed to be equal to the OCV.

\subsection{One-Dimensional Single-Particle Model}

The direct-pulse fitting method relies on a one-dimensional electrochemical model, where the surface lithium flux is coupled to the solid voltage through the Butler-Volmer equation (\cref{eq:BV}). Inside the particles, lithium transport is governed by the one-dimensional, differential, conservation equation in the radial direction $r$:
\begin{equation}
    \frac{\partial C_\mathrm{Li}}{\partial t} = -\frac{1}{r^2}\frac{\partial}{\partial r}\left(r^2 j\right),
    \label{eq:diffeq}
\end{equation}
where the lithium flux $j$ can be assumed to be either ideal:
\begin{equation}
    j = -D_\mathrm{Li}\frac{\partial C_\mathrm{Li}}{\partial r}
    \label{eq:ideal}
\end{equation}
or non-ideal:
\begin{equation}
    j = \frac{D_\mathrm{Li} F C_\mathrm{Li}}{R T}\frac{\partial V_\mathrm{Eq}}{\partial r}.
    \label{eq:nonideal}
\end{equation}
This constitutes a single, partial-differential equation which can be solved when subject to the specified flux at the particle surface and a symmetry boundary condition applied at the particle center. Initially, the lithium concentration in the particles is assumed to be constant. A summary of the parameters involved in the model is shown in \cref{tab:params}. The reaction rate constant $k$ was evaluated independently for each step from the IR drop, which we assumed to be the initial change in voltage over the first 2 seconds. We tested this assumption by comparing cutoffs of 0.1 and 5 seconds for the IR drop and obtained an average change in the diffusion coefficients of 13\% despite an average change of 94\% in the reaction rate constants. During this drop, the lithium concentrations and equilibrium potential are assumed to be constant, and the current is instantly changed. As such, one may calculate $k$ explicitly by rearranging \cref{eq:BV,eq:ECD}. There are several assumptions that go into the above formulation: that all particles are assumed to be spherical, that no lithium gradients are present in the electrolyte, and that all side reactions have been neglected.

\begin{table*}
    \centering
    \caption{Relevant parameters used in the model fitting and predictions.}
    \label{tab:params}
    \begin{tabular}{@{}lll@{}}
        \toprule
        Parameter & Description & Value \\ \midrule
        $F$ & Faraday constant & \SI{96485}{C/mol} \\
        $\alpha$ & Charge transfer coefficient & 0.5 \\
        $R$ & Ideal gas constant & \SI{8.314}{J/(K.mol)} \\
        $T$ & Temperature & \SI{298}{K} \\
        $C_\mathrm{Li^+}$ & Electrolyte concentration & \SI{1000}{mol/m^3} \\
        $C_\mathrm{max,Li}$ & Maximum lithium concentration & \SI{80000}{mol/m^3} \\
        $c_\mathrm{max}$ & Theoretical capacity & \SI{894}{mA.h/g} \\
        $x_0$ & Initial x in \ce{Li_{x}FeS2} & 1.2 \\
        $R_p$ & Particle radius (unless otherwise specified) & \SI{1.045}{\micro m} \\
        \bottomrule
    \end{tabular}
\end{table*}

\subsection{Fitting Protocols}

A comparison of all fitting methods used throughout this work is provided in \cref{tab:fitting}. The square root fit used throughout this work was originally proposed by \citet{Weppner1977}. Through approximations to the solution of the ideal diffusion equation during charge/discharge, the diffusion coefficient may be evaluated algebraically from the slope of the potential as a function of the square root of the step time (\cref{eq:D_sqrt}). This behavior is constrained to times much shorter than the effective diffusion time $R_p^2/D_\mathrm{Li}$. However, the fit data must be far enough from the step change to neglect the ohmic overpotential, the kinetic overpotential, and behavior that depends on the concentration profile from the previous step. In this work, the diffusion coefficients obtained through this method were obtained from measurements during the discharge steps after 50 seconds. The active surface area was evaluated based on the assumed particle radius (\SI{1.045}{\micro\meter} unless otherwise stated); the current $i_\mathrm{Tot}$ was determined based on the applied C-rate and the theoretical max capacity; and the derivative of the equilibrium potential $d V_\mathrm{Eq}/d C_\mathrm{Li}$ was determined from a high-order polynomial fit to the experimentally determined OCV data. Note that the value for the assumed particle radius is derived from the distribution shown in \cref{fig:materials:PSD} as three times the integral of the weighted volume divided by the integral of the weighted surface area.

\begin{table*}
    \centering
    \small
    \caption{Comparison of different GITT diffusion coefficient estimation methods used throughout this work.}
    \label{tab:fitting}
    {\renewcommand{\arraystretch}{1.2}
    \begin{tabular}{@{}llp{0.25\linewidth}p{0.25\linewidth}@{}}
        \toprule
         & Exponential & Square Root & Direct-Pulse Fitting \\ \midrule
        Part of GITT Pulse & Rest & Charge/discharge & Full pulse \\
        Diffusion Model & Ideal & Ideal & Ideal or non-ideal \\
        Governing Equations & \cref{eq:D_expo} & \cref{eq:D_sqrt} & \cref{eq:BV,eq:ECD,eq:diffeq,eq:ideal,eq:nonideal} \\
        Fitting Method & Linear regression & Linear regression & Non-linear regression \\
        Key Assumptions & Close to OCV & Step time is less than diffusion time & Negligible gradients in electrolyte \\
        \bottomrule
    \end{tabular}}
\end{table*}

During the rest step, the surface lithium concentration for a particle will obey an exponential relaxation at long times according to a simplified, first-order solution to the ideal diffusion equation. For a spherical particle subjected to the boundary conditions discussed in the preceding section, the surface lithium concentration may be converted to an overpotential $\eta$ which will obey:
\begin{equation}
    \frac{d \ln\left(\eta\right)}{d t} = -\frac{20.2 D_\mathrm{Li}}{R_p^2}.
    \label{eq:D_expo}
\end{equation}
Here, the factor of 20.2 is derived from the first eigenvalue to the ideal diffusion equation in spherical coordinates. Note that this representation requires two assumptions: that this behavior is only applicable at times far from the step change in applied current, and (as the direct connection between the overpotential and the surface lithium concentration is unknown) that the solid voltage is close to the OCV. To determine diffusion coefficients from this method, only time points after the first hour of resting were considered. Using the natural logarithm of these points, the overpotential was fit to a linear regression model. As the OCV is not directly known, this single parameter for each rest step is left as a nonlinear parameter to be fit as discussed in the OCV Evaluation section above.

For the direct-pulse fitting method, \cref{eq:diffeq} was solved using MATLAB’s built in function, \texttt{pdepe}. Only steps within the known OCV were considered. For the individual pulse fits, the model was fit to both the discharge and rest step, and the sum of the squared voltage residuals, normalized by the number of total data points, was used as the objective function. For the full-curve fits, the model was fit to all steps in the considered range with the normalized sum of the squared voltage residuals used as the objective function. The fitting was performed with MATLAB’s \texttt{fminsearch} to evaluate the single nonlinear parameter $D_\mathrm{Li}$. For all predictions, the same partial differential equation was used with a single diffusion coefficient, corresponding to the optimal full-curve coefficient for the direct fitting methods and the median diffusion coefficient value for the algebraic methods.

\section{Results and Discussion}
\subsection{Single-Particle Diffusion Coefficient Evaluation}

\begin{figure*}
    \includegraphics[width=\linewidth]{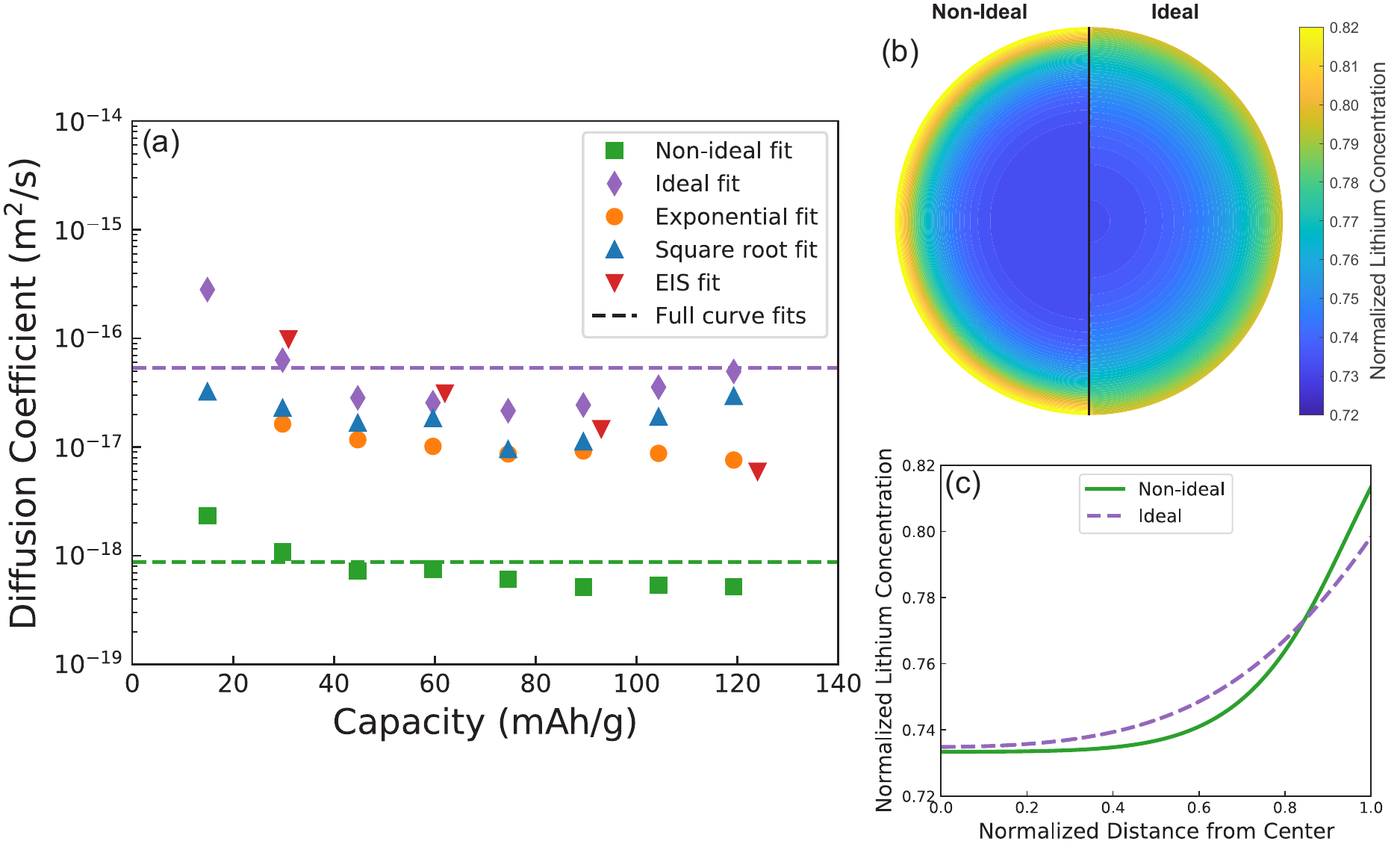}
    {
        \phantomsubcaption\label{fig:singleparticle:diffcoefs}
        \phantomsubcaption\label{fig:singleparticle:Licontours}
        \phantomsubcaption\label{fig:singleparticle:lithprof}
    }
    \caption{\textbf{Diffusion coefficient comparison and lithium profiles.} (a) Comparison of diffusion coefficients calculated using various methods as a function of specific capacity. Points represent individual pulse fits and dashed lines represent full-curve fits. (b) (left) Non-ideal and (right) ideal lithium contour comparison at the end of a discharge step at $\sim$\SI{1.9}{V}. (c) Lithium profile as a function of distance from particle center at the end of a discharge step at $\sim$\SI{1.9}{V}.}
    \label{fig:singleparticle}
\end{figure*}

In \cref{fig:singleparticle:diffcoefs}, we compare the diffusion coefficients obtained from all GITT analysis methods, as well as diffusion coefficients evaluated from EIS measurements (with details of the EIS measurements provided in the Supporting Information); we also include the diffusion coefficients obtained through the entire GITT curve fitting for ideal and non-ideal solution theory as dashed lines. All diffusion coefficients shown in \cref{fig:singleparticle:diffcoefs} are at room temperature, but additional measurements and diffusion coefficients at \SI{60}{\celsius} are presented in the Supporting Information. A succinct comparison of all methods considered may be found in \cref{tab:fitting} in the Experimental Section.

For all methods that assume an ideal solution (i.e. ideal direct-pulse fitting, exponential fitting, and square root fitting), the diffusion coefficients come within an order of magnitude at mid-range capacities. However, direct-pulse fitting with ideal solution theory consistently results in higher diffusion coefficient values than those obtained through the exponential or square root fit -- likely artifacts of the inherent assumptions associated with the exponential fit and square root fit methods. Diffusion coefficients evaluated through the ideal direct-pulse fitting method and square root method also exhibit a minimum at middle capacities. We hypothesize that the minimum is due to limitations of the ideal diffusion model which should deviate most significantly from the non-ideal diffusion model when the equilibrium potential is changing rapidly (i.e. at the beginning and end of discharge). However, it is worth noting that this behavior has been seen in other materials and warrants future investigation to determine whether this is an intrinsic property of intercalation materials or a fitting artifact\cite{Amin2016,Amin2015,Noh2013}. EIS-determined diffusion coefficients also fall within an order of magnitude of those evaluated through GITT using ideal solution theory. However, considerably more variation exists across EIS-determined diffusion coefficients, which likely corresponds to the limitations and uncertainties of EIS as discussed in the Supporting Information. The discrepancy in ideal diffusion coefficients demonstrates the high variability in obtained diffusion coefficient values depending on which method is used to extract them.

The discrepancy between diffusion coefficients is more notable when comparing ideal solution theory with non-ideal solution theory. Diffusion coefficients obtained for direct-pulse fitting using non-ideal solution theory are more than an order of magnitude less than those obtained through direct fitting using ideal solution theory, which suggests a fundamental difference between ideal and non-ideal solution models. The non-ideal diffusion coefficients are much lower than the ideal diffusion coefficients because the magnitude of the excess chemical potential gradient in the particles is larger than the magnitude of the ideal chemical potential gradient. Thus, to counterbalance this enhanced transport provided by the non-ideal diffusion model, the corresponding diffusion coefficients must be less than the ideal case. The observation of such a significant decrease in the diffusion coefficients when using non-ideal solution theory lends credibility to the assertion that \ce{FeS2} is a highly non-ideal material. It also highlights that the diffusion coefficient is simply a model parameter, and as such has both different meanings and values in different models.

To further understand this fundamental difference, the lithium concentration profiles at the end of a discharge step are shown in \cref{fig:singleparticle:Licontours,fig:singleparticle:lithprof} for a single diffusion coefficient used for the entire GITT curve. Qualitatively, non-ideal solution theory predicts a steeper near-surface lithium gradient than ideal solution theory, likely as a result of the difference in diffusion coefficient variation across the full GITT curve. The ideal diffusion coefficients have more variation than non-ideal diffusion coefficients, particularly at low capacities, and a single, ideal diffusion coefficient will not be able to represent the full curve as accurately as a single non-ideal diffusion coefficient due to this variability. At mid-range capacities, which correspond to the specific pulse depicted in \cref{fig:singleparticle:Licontours,fig:singleparticle:lithprof}, the full-curve, ideal diffusion coefficient is higher than the ideal diffusion coefficient for that individual pulse; and thus, the predicted lithium profiles exhibit faster diffusion than the non-ideal case, which is expected to be more accurate, as the full-curve, non-ideal diffusion coefficient value is closer to the value for the individual pulse.

\subsection{Extension to Polydisperse Electrodes}

Existing approaches for evaluating diffusion coefficients were derived for a single-particle system; however, a single-particle approach requires accurate knowledge of the effective particle radius or the active surface area for the system\cite{Chouchane2020}, which are difficult to estimate because particle radii can vary significantly throughout an electrode. The direct-pulse fitting method can be extended for any number of polydisperse particles, due to the quadratic form of the Butler-Volmer equation which can be solved for the solid-phase voltage:
\begin{equation} \label{eq:vs}
    \begin{split}
        V_S &= \frac{RT}{\alpha F}\ln\left(\frac{-b+\sqrt{b^2-4ac}}{2a}\right),  \\
        a &= \sum_{p=1}^{n_p}A_p i_{0,p}\exp\left(-\frac{\alpha F V_{\mathrm{Eq},p}}{RT}\right), \\
        b &= -i_\mathrm{Tot}, \\
        c &= -\sum_{p=1}^{n_p}A_p i_{0,p}\exp\left(\frac{\alpha F V_{\mathrm{Eq},p}}{RT}\right).
    \end{split} 
\end{equation}
Here, $n_p$ is the number of particles, $A_p$ is the active surface area for particle $p$, $i_{0,p}$ is the exchange current density for particle $p$, and $i_\mathrm{Tot}$ is the total applied current. The solid-phase voltage is assumed to be constant for all particles due to the conductive additive in the binder. Once the solid phase voltage is determined, the flux for each particle may be calculated explicitly. By using a multiparticle electrochemical model, we can more accurately represent the electrode and clarify its effective particle radius.

\begin{figure*}
    \includegraphics[width=\linewidth]{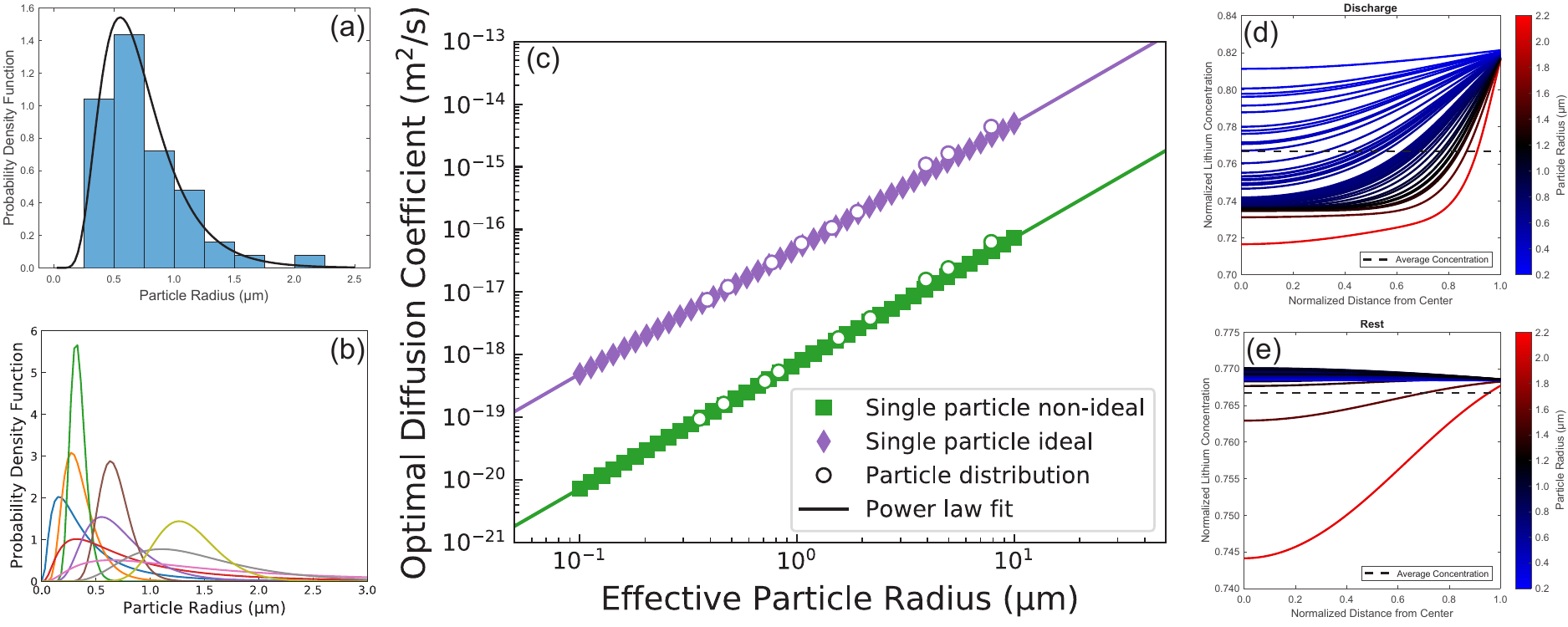}
    {
        \phantomsubcaption\label{fig:multiparticle:PDF}
        \phantomsubcaption\label{fig:multiparticle:PDFs}
        \phantomsubcaption\label{fig:multiparticle:difftime}
        \phantomsubcaption\label{fig:multiparticle:discharge}
        \phantomsubcaption\label{fig:multiparticle:rest}
    }
    \caption{\textbf{Summary of results for polydisperse electrodes.} (a) A histogram of 50 different particles randomly selected from the log-normal distribution corresponding to the ball-milled \ce{FeS2}. (b) Various other log-normal probability distributions considered with varying mean and standard deviation. (c) Calculated diffusion coefficients for (filled symbols) single-particle fits and (open symbols) particle distribution fits, where lines represent power law fits. (d,e) Lithium profiles in all 50 particles as a function of normalized distance from center at the end of a (d) discharge and (e) rest step at $\sim$\SI{1.9}{V} and $\sim$\SI{2.0}{V}, respectively. Colors represent particle size, corresponding to the scale on the right; and the black, dashed line represents the volume average concentration for all particles.}
    \label{fig:multiparticle}
\end{figure*}

To better determine the role of polydispersity, we simulate several log-normal distributions of particle radii (\cref{fig:multiparticle:PDF,fig:multiparticle:PDFs}), obtaining a diffusion coefficient from each simulation (\cref{fig:multiparticle:difftime}). As the radius for each particle in the simulation is unique, we use an effective particle radius, defined as $R_{p,\mathrm{Eff}}=3V_\mathrm{Tot}/A_\mathrm{Tot}$ (i.e. equal specific surface area to the collection of particles), where the total volume and surface area are determined from the simulated collection of particles. Using this methodology, we are able to compare with single-particle simulations to obtain a collapse of the calculated diffusion coefficients for single-particle and polydisperse systems, as shown in \cref{fig:multiparticle:difftime}\cite{Verma2017}. The collapse demonstrates that diffusion coefficients scale with respect to the effective particle radius squared or, equivalently, that the effective diffusion time $R_p^2/D_\mathrm{Li}$ is constant regardless of the assumed radius. Moreover, for our considered distributions, a multiparticle model behaves similarly to a single-particle model, with effective particle radius as defined above. The lithium concentration profiles for all simulated particles in a particular distribution are shown in \cref{fig:multiparticle:discharge,fig:multiparticle:rest} at the end of a discharge step and the end of the subsequent rest step. During discharge, the small particles in the distribution over-lithiate, then delithiate in the subsequent rest step as the lithium moves to fill the larger particles, indicating that electrochemical reactions are still occurring even in the absence of applied current. Additionally, at the end of the 4-hour rest step, one can see that the distribution is not yet in equilibrium, and lithium is still moving throughout the system, which indicates that the voltage has not reached the OCV for a homogeneous system. For a single-particle system, this is analogous to the lithium profile not being constant throughout a particle at the end of a rest step.

\subsection{Validation through Discharge Predictions}

\begin{figure*}
    \includegraphics[width=\linewidth]{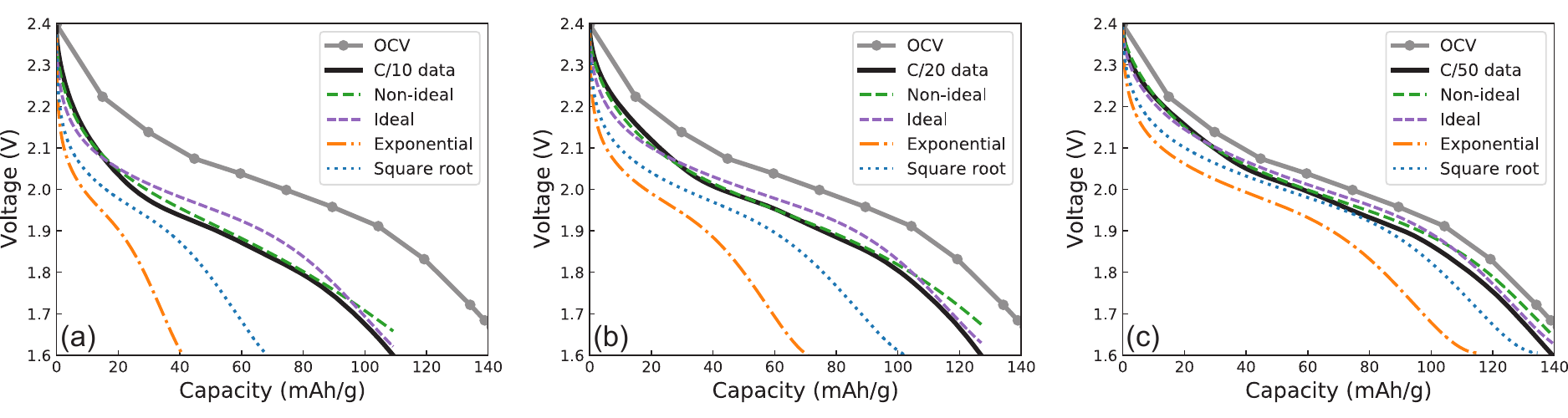}
    {
        \phantomsubcaption\label{fig:validation:C10}
        \phantomsubcaption\label{fig:validation:C20}
        \phantomsubcaption\label{fig:validation:C50}
    }
    \caption{\textbf{Discharge profiles at different discharge rates.} (black lines) Experimental discharge data and (colored, dashed lines) one-dimensional predictions based on diffusion coefficients obtained through different methods. Results for (a) C/10, (b) C/20, and (c) C/50 are shown, and OCV is shown on each for reference.}
    \label{fig:validation}
\end{figure*}

To validate our direct-pulse fitting method, we collected potential measurements during discharge at three different discharge rates (C/10, C/20, and C/50, where C is the rate required to discharge the entire electrode in one hour). Using the same single-particle model used for extracting the diffusion coefficients, we predict the response of the cell for all previously obtained diffusion coefficient values (\cref{fig:validation}). Note that we use the ideal diffusion model for ideal, exponential, and square root predictions, where the only difference is the diffusion coefficient value; the non-ideal prediction uses the non-ideal diffusion model. The diffusion coefficients obtained from our direct-pulse fitting methods predict the experimental cycling data two orders of magnitude better than the diffusion coefficients obtained through the square root fit and exponential fit methods. Moreover, among the direct-pulse fitting methods, non-ideal solution theory provides the best overall prediction to the experimental data, most notably at higher C-rates (\cref{fig:validation:C10}).

Despite the diffusion coefficient being lower for the non-ideal case, the non-ideal prediction agrees best with the experimental data. No loss of agreement is observed despite lower diffusion coefficient value because the non-ideal diffusion model offers enhanced transport through the potential gradient driving force when compared to the ideal diffusion model. The non-ideality of lithium insertion into \ce{FeS2} has been previously reported\cite{LEMEHAUTE1981}, and our results, particularly at faster rates than considered in the GITT measurements (\cref{fig:validation:C10}), further demonstrate the need to use non-ideal solution theory for lithium transport within the active material to properly represent the experimental measurements. Our predictions are especially promising given the simplicity of our current model, which highlight that solid-state lithium diffusion is the leading loss mechanism, although further refinements may be made by incorporating the binder domain, electronic transport equations, accurate particle morphologies, or mechanics.

\subsection{Application to Other Cathode Materials}

\begin{figure*}
    \centering
    \includegraphics[width=0.6\linewidth]{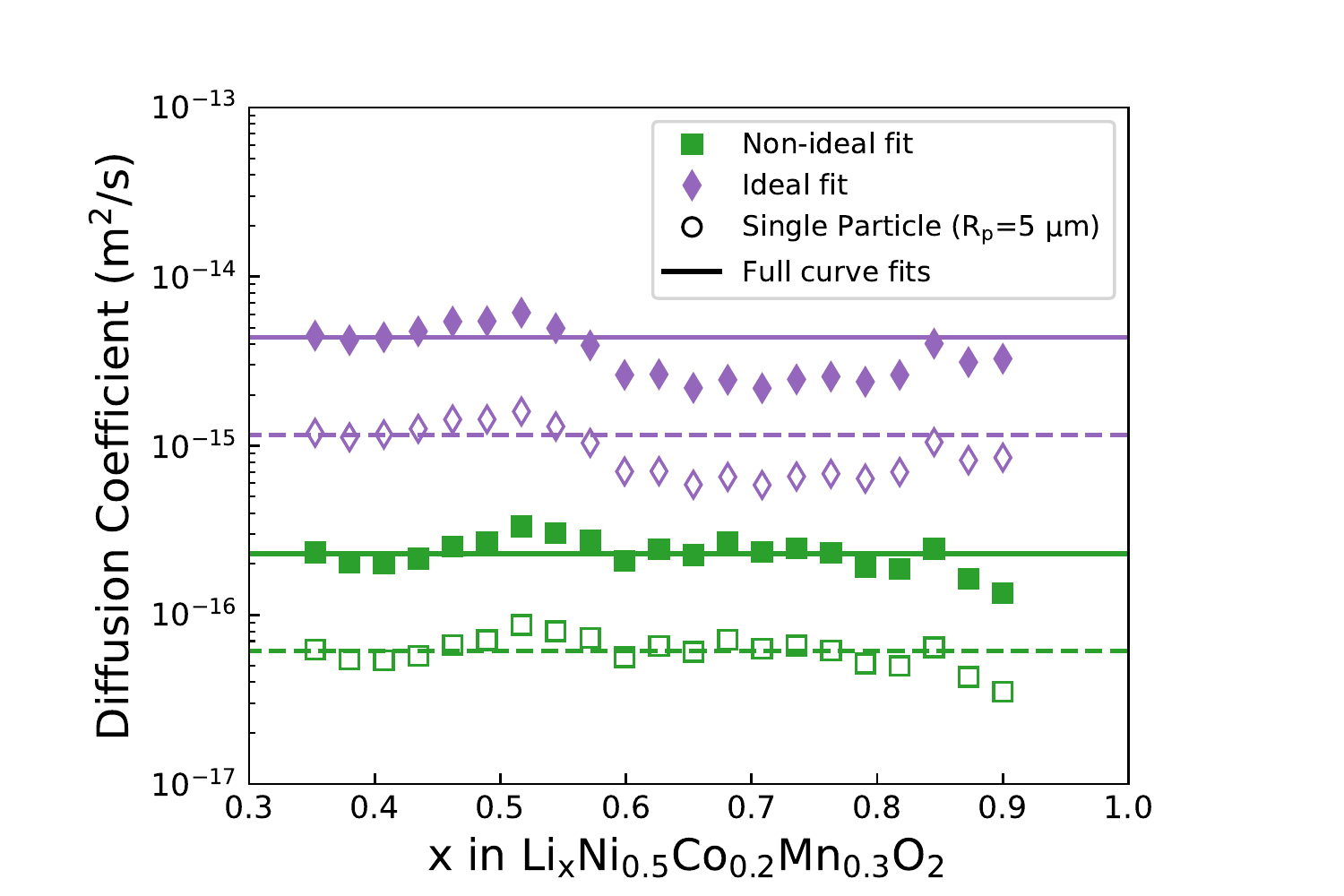}
    \caption{\textbf{NCM523 diffusion coefficients.} Comparison of diffusion coefficients evaluated through the direct-pulse fitting method. Filled points correspond to a multiparticle simulation based on the distribution function provided in \citet{Nickol2020}. Open points correspond to results for a single particle with a radius $R_p$ of \SI{5}{\micro m}. Points represent individual pulse fits, and lines represent full-curve fits. The solid lines correspond to a distribution of particles and the dashed lines correspond to a single particle with a \SI{5}{\micro m} radius.}
    \label{fig:NMC}
\end{figure*}

Our technique for analyzing GITT data can be similarly applied to any intercalating cathode material. For example, we applied this technique to recently published GITT data for NCM523\cite{Nickol2020} and used the provided particle size distribution and charging data at \SI{30}{\celsius} to predict diffusion coefficients. As shown in \cref{fig:NMC}, obtained values spanned more than two orders of magnitude depending on which model and assumed particle radius was used. Specifically, we obtained values of \SI{4e-15}{\meter^2\per\second} and \SI{2e-16}{m^2/s} for the full-curve fit using ideal and non-ideal solution theory, respectively. The ideal diffusion coefficient obtained through direct fitting is approximately four times larger than the highest reported values in the original paper\cite{Nickol2020}. To determine whether this discrepancy should be attributed to different assumed active surface areas or differences between the fitting methods, we also fit diffusion coefficients for a single particle with radius \SI{5}{\micro m} (as in the original NCM523 calculations). The resulting non-ideal diffusion coefficient (\SI{6e-17}{m^2/s}) is significantly lower than any value obtained in the original paper, whereas the ideal diffusion coefficient (\SI{1e-15}{m^2/s}) is similar to the largest values obtained in the original paper\cite{Nickol2020}. Details on the calculations for NCM523 are provided in the Supporting Information.

Unfortunately, cycling data were not provided in the original paper, so the accuracy of all diffusion coefficients cannot be independently tested and further investigation is warranted. However, the large discrepancy across diffusion coefficients, particularly comparing ideal and non-ideal diffusion coefficients, agree with our observations for \ce{FeS2} diffusion coefficients. By applying our GITT analytical techniques to NCM523 in addition to \ce{FeS2} cathodes, we show the generality of the approach and demonstrate the discrepencies that arise depending on the analytical method used. Thus, when using diffusion coefficients from literature, one should carefully identify how the active surface area was measured and whether the diffusion was assumed to be ideal or non-ideal, as these assumptions can significantly affect the derived diffusion coefficients.

\section{Conclusions}

Throughout this work, we demonstrated that diffusion coefficients obtained through GITT data may vary significantly, depending on which method and assumed active surface area is used. To address the latter, we conducted multiparticle simulations and demonstrated that particle distributions behave similarly to a single particle with a radius proportional to the total volume-to-surface-area ratio. By using the correct radius and directly fitting the GITT response to obtain the diffusion coefficient, we improved the accuracy of discharge predictions at multiple discharge rates by two orders of magnitude. Furthermore, in testing both non-ideal and ideal diffusion models, we found that the required diffusion coefficients for non-ideal solution theory are more than an order of magnitude lower than those required for ideal solution theory. Using the optimal values for each model, non-ideal solution theory was shown to predict the experimental data significantly better, most notably at high discharge rates.

This work demonstrates that GITT is an effective tool for evaluating diffusion coefficients but that the existing methods for analyzing GITT data are flawed. When interpreting data and analysis from previous GITT studies on active materials, it is particularly important to be wary of how the active surface area was determined, whether the solution was assumed to be ideal, and what assumptions underlie the analysis method used to extract the diffusion coefficients. Our results also indicate that diffusion coefficients based on non-ideal solution theory are not the same as those obtained for ideal assumptions. Thus, one should be careful not to use ideal diffusion coefficients in a non-ideal model. Additionally, algebraic methods derived from simplified solutions to the governing diffusion equation typically provide poor estimates of the intrinsic diffusion coefficient of the material which can be clearly seen through subsequent predictions. Moving beyond electrochemical systems, the underlying theory of evaluating transient properties of a system based on alternating load and rest steps is employed in fields ranging from the mechanics of complex materials to the response monitoring of biological systems. The concepts presented in this work suggest that simplified solutions to these responses should be rigorously tested against more empirical models. Despite the promising results of the direct-fitting method coupled with non-ideal solution theory, there remain several avenues for improvement in modeling and simulation of electrochemical systems. Particularly, our model is designed for intercalation reactions, and with increasing interest in next-generation batteries, it will be essential to develop models and experimental methods to measure physical properties for conversion reactions. Going forward, it will also be important to systematically study the impact that the binder domain, electronic conductivity, electrolyte, particle morphology, and system mechanics have on the overall cell behavior during discharge and GITT measurements.

\section{Author Contributions}

J. S. H. and S. A. R. performed the modeling and designed the study; G. W. and B. S. D. conducted the GITT and cycling experiments; D. S. A. and A. A. T. conducted the EIS experiments and interpretation; I. V. K. and T. N. L. performed the laser diffraction analysis; and J. S. H., D. S. A., G. W., and I. V. K. all contributed to writing the manuscript with input from all authors.

\begin{acknowledgement}

This paper describes objective technical results and analysis. Any subjective views or opinions that might be expressed in the paper do not necessarily represent the views of the U.S. Department of Energy or the United States Government. Supported by the Laboratory Directed Research and Development program at Sandia National Laboratories, a multimission laboratory managed and operated by National Technology and Engineering Solutions of Sandia, LLC., a wholly owned subsidiary of Honeywell International, Inc., for the U.S. Department of Energy's National Nuclear Security Administration under contract DE-NA-0003525. The authors thank Katharine Harrison for her guidance and feedback on the project, Jacquilyn Weeks for her help with editing the manuscript, and Alexander Nickol for providing the published NCM523 data. We also appreciate pre-submission peer review from Daniel Wesolowski and Eric Allcorn.

\end{acknowledgement}

\begin{suppinfo}

Derivation for the exponential fit method used throughout this work; \SI{60}{\celsius} GITT measurements and corresponding diffusion coefficients; EIS measurements and discussion; and technical details for NCM523 diffusion coefficient extraction methods.

\end{suppinfo}

\bibliography{gitt}

\providecommand{\latin}[1]{#1}
\makeatletter
\providecommand{\doi}
  {\begingroup\let\do\@makeother\dospecials
  \catcode`\{=1 \catcode`\}=2 \doi@aux}
\providecommand{\doi@aux}[1]{\endgroup\texttt{#1}}
\makeatother
\providecommand*\mcitethebibliography{\thebibliography}
\csname @ifundefined\endcsname{endmcitethebibliography}
  {\let\endmcitethebibliography\endthebibliography}{}
\begin{mcitethebibliography}{33}
\providecommand*\natexlab[1]{#1}
\providecommand*\mciteSetBstSublistMode[1]{}
\providecommand*\mciteSetBstMaxWidthForm[2]{}
\providecommand*\mciteBstWouldAddEndPuncttrue
  {\def\EndOfBibitem{\unskip.}}
\providecommand*\mciteBstWouldAddEndPunctfalse
  {\let\EndOfBibitem\relax}
\providecommand*\mciteSetBstMidEndSepPunct[3]{}
\providecommand*\mciteSetBstSublistLabelBeginEnd[3]{}
\providecommand*\EndOfBibitem{}
\mciteSetBstSublistMode{f}
\mciteSetBstMaxWidthForm{subitem}{(\alph{mcitesubitemcount})}
\mciteSetBstSublistLabelBeginEnd
  {\mcitemaxwidthsubitemform\space}
  {\relax}
  {\relax}

\bibitem[Zhang(2019)]{Zhang2019}
Zhang,~S.~S. Identifying rate limitation and a guide to design of fast-charging
  Li-ion battery. \emph{{InfoMat}} \textbf{2019}, \emph{2}, 942--949\relax
\mciteBstWouldAddEndPuncttrue
\mciteSetBstMidEndSepPunct{\mcitedefaultmidpunct}
{\mcitedefaultendpunct}{\mcitedefaultseppunct}\relax
\EndOfBibitem
\bibitem[Rodrigues \latin{et~al.}(2017)Rodrigues, Babu, Gullapalli, Kalaga,
  Sayed, Kato, Joyner, and Ajayan]{Rodrigues2017}
Rodrigues,~M.-T.~F.; Babu,~G.; Gullapalli,~H.; Kalaga,~K.; Sayed,~F.~N.;
  Kato,~K.; Joyner,~J.; Ajayan,~P.~M. A materials perspective on Li-ion
  batteries at extreme temperatures. \emph{Nature Energy} \textbf{2017},
  \emph{2}, 1--14\relax
\mciteBstWouldAddEndPuncttrue
\mciteSetBstMidEndSepPunct{\mcitedefaultmidpunct}
{\mcitedefaultendpunct}{\mcitedefaultseppunct}\relax
\EndOfBibitem
\bibitem[Zeng \latin{et~al.}(2019)Zeng, Li, El-Hady, Alshitari, Al-Bogami, Lu,
  and Amine]{Zeng2019}
Zeng,~X.; Li,~M.; El-Hady,~D.~A.; Alshitari,~W.; Al-Bogami,~A.~S.; Lu,~J.;
  Amine,~K. Commercialization of Lithium Battery Technologies for Electric
  Vehicles. \emph{Advanced Energy Materials} \textbf{2019}, \emph{9},
  1900161\relax
\mciteBstWouldAddEndPuncttrue
\mciteSetBstMidEndSepPunct{\mcitedefaultmidpunct}
{\mcitedefaultendpunct}{\mcitedefaultseppunct}\relax
\EndOfBibitem
\bibitem[Ivanishchev \latin{et~al.}(2017)Ivanishchev, Ushakov, Ivanishcheva,
  Churikov, Mironov, Fedotov, Khasanova, and Antipov]{Ivanishchev2017}
Ivanishchev,~A.~V.; Ushakov,~A.~V.; Ivanishcheva,~I.~A.; Churikov,~A.~V.;
  Mironov,~A.~V.; Fedotov,~S.~S.; Khasanova,~N.~R.; Antipov,~E.~V. Structural
  and electrochemical study of fast Li diffusion in Li3V2({PO}4)3-based
  electrode material. \emph{Electrochimica Acta} \textbf{2017}, \emph{230},
  479--491\relax
\mciteBstWouldAddEndPuncttrue
\mciteSetBstMidEndSepPunct{\mcitedefaultmidpunct}
{\mcitedefaultendpunct}{\mcitedefaultseppunct}\relax
\EndOfBibitem
\bibitem[Weppner and Huggins(1977)Weppner, and Huggins]{Weppner1977}
Weppner,~W.; Huggins,~R.~A. Determination of the Kinetic Parameters of
  Mixed-Conducting Electrodes and Application to the System Li3Sb.
  \emph{Journal of The Electrochemical Society} \textbf{1977}, \emph{124},
  1569--1578\relax
\mciteBstWouldAddEndPuncttrue
\mciteSetBstMidEndSepPunct{\mcitedefaultmidpunct}
{\mcitedefaultendpunct}{\mcitedefaultseppunct}\relax
\EndOfBibitem
\bibitem[Verma \latin{et~al.}(2017)Verma, Smith, Santhanagopalan, Abraham, Yao,
  and Mukherjee]{Verma2017}
Verma,~A.; Smith,~K.; Santhanagopalan,~S.; Abraham,~D.; Yao,~K.~P.;
  Mukherjee,~P.~P. Galvanostatic Intermittent Titration and Performance Based
  Analysis of {LiNi}0.5Co0.2Mn0.3O2Cathode. \emph{Journal of The
  Electrochemical Society} \textbf{2017}, \emph{164}, A3380--A3392\relax
\mciteBstWouldAddEndPuncttrue
\mciteSetBstMidEndSepPunct{\mcitedefaultmidpunct}
{\mcitedefaultendpunct}{\mcitedefaultseppunct}\relax
\EndOfBibitem
\bibitem[Nickol \latin{et~al.}(2020)Nickol, Schied, Heubner, Schneider,
  Michaelis, Bobeth, and Cuniberti]{Nickol2020}
Nickol,~A.; Schied,~T.; Heubner,~C.; Schneider,~M.; Michaelis,~A.; Bobeth,~M.;
  Cuniberti,~G. {GITT} Analysis of Lithium Insertion Cathodes for Determining
  the Lithium Diffusion Coefficient at Low Temperature: Challenges and
  Pitfalls. \emph{Journal of The Electrochemical Society} \textbf{2020},
  \emph{167}, 090546\relax
\mciteBstWouldAddEndPuncttrue
\mciteSetBstMidEndSepPunct{\mcitedefaultmidpunct}
{\mcitedefaultendpunct}{\mcitedefaultseppunct}\relax
\EndOfBibitem
\bibitem[Deng and Lu(2020)Deng, and Lu]{Deng2020}
Deng,~C.; Lu,~W. Consistent diffusivity measurement between Galvanostatic
  Intermittent Titration Technique and Electrochemical Impedance Spectroscopy.
  \emph{Journal of Power Sources} \textbf{2020}, \emph{473}, 228613\relax
\mciteBstWouldAddEndPuncttrue
\mciteSetBstMidEndSepPunct{\mcitedefaultmidpunct}
{\mcitedefaultendpunct}{\mcitedefaultseppunct}\relax
\EndOfBibitem
\bibitem[Finegan \latin{et~al.}(2021)Finegan, Zhu, Feng, Keyser, Ulmefors, Li,
  Bazant, and Cooper]{Finegan2021}
Finegan,~D.~P.; Zhu,~J.; Feng,~X.; Keyser,~M.; Ulmefors,~M.; Li,~W.;
  Bazant,~M.~Z.; Cooper,~S.~J. The Application of Data-Driven Methods and
  Physics-Based Learning for Improving Battery Safety. \emph{Joule}
  \textbf{2021}, \emph{5}, 316--329\relax
\mciteBstWouldAddEndPuncttrue
\mciteSetBstMidEndSepPunct{\mcitedefaultmidpunct}
{\mcitedefaultendpunct}{\mcitedefaultseppunct}\relax
\EndOfBibitem
\bibitem[Lu \latin{et~al.}(2020)Lu, Bertei, Finegan, Tan, Daemi, Weaving,
  O'Regan, Heenan, Hinds, Kendrick, Brett, and Shearing]{Lu2020}
Lu,~X.; Bertei,~A.; Finegan,~D.~P.; Tan,~C.; Daemi,~S.~R.; Weaving,~J.~S.;
  O'Regan,~K.~B.; Heenan,~T. M.~M.; Hinds,~G.; Kendrick,~E.; Brett,~D. J.~L.;
  Shearing,~P.~R. 3D microstructure design of lithium-ion battery electrodes
  assisted by X-ray nano-computed tomography and modelling. \emph{Nature
  Communications} \textbf{2020}, \emph{11}, 1--13\relax
\mciteBstWouldAddEndPuncttrue
\mciteSetBstMidEndSepPunct{\mcitedefaultmidpunct}
{\mcitedefaultendpunct}{\mcitedefaultseppunct}\relax
\EndOfBibitem
\bibitem[Mistry \latin{et~al.}(2021)Mistry, Franco, Cooper, Roberts, and
  Viswanathan]{Mistry2021}
Mistry,~A.; Franco,~A.~A.; Cooper,~S.~J.; Roberts,~S.~A.; Viswanathan,~V. How
  Machine Learning Will Revolutionize Electrochemical Sciences. \emph{{ACS}
  Energy Letters} \textbf{2021}, 1422--1431\relax
\mciteBstWouldAddEndPuncttrue
\mciteSetBstMidEndSepPunct{\mcitedefaultmidpunct}
{\mcitedefaultendpunct}{\mcitedefaultseppunct}\relax
\EndOfBibitem
\bibitem[Bielefeld \latin{et~al.}(2020)Bielefeld, Weber, and
  Janek]{Bielefeld2020}
Bielefeld,~A.; Weber,~D.~A.; Janek,~J. Modeling Effective Ionic Conductivity
  and Binder Influence in Composite Cathodes for All-Solid-State Batteries.
  \emph{{ACS} Applied Materials {\&} Interfaces} \textbf{2020}, \emph{12},
  12821--12833\relax
\mciteBstWouldAddEndPuncttrue
\mciteSetBstMidEndSepPunct{\mcitedefaultmidpunct}
{\mcitedefaultendpunct}{\mcitedefaultseppunct}\relax
\EndOfBibitem
\bibitem[Ferraro \latin{et~al.}(2020)Ferraro, Trembacki, Brunini, Noble, and
  Roberts]{Ferraro2020}
Ferraro,~M.~E.; Trembacki,~B.~L.; Brunini,~V.~E.; Noble,~D.~R.; Roberts,~S.~A.
  Electrode Mesoscale as a Collection of Particles: Coupled Electrochemical and
  Mechanical Analysis of {NMC} Cathodes. \emph{Journal of The Electrochemical
  Society} \textbf{2020}, \emph{167}, 013543\relax
\mciteBstWouldAddEndPuncttrue
\mciteSetBstMidEndSepPunct{\mcitedefaultmidpunct}
{\mcitedefaultendpunct}{\mcitedefaultseppunct}\relax
\EndOfBibitem
\bibitem[Delacourt \latin{et~al.}(2011)Delacourt, Ati, and
  Tarascon]{Delacourt2011}
Delacourt,~C.; Ati,~M.; Tarascon,~J.~M. Measurement of Lithium Diffusion
  Coefficient in {LiyFeSO}4F. \emph{Journal of The Electrochemical Society}
  \textbf{2011}, \emph{158}, A741\relax
\mciteBstWouldAddEndPuncttrue
\mciteSetBstMidEndSepPunct{\mcitedefaultmidpunct}
{\mcitedefaultendpunct}{\mcitedefaultseppunct}\relax
\EndOfBibitem
\bibitem[Amin and Chiang(2016)Amin, and Chiang]{Amin2016}
Amin,~R.; Chiang,~Y.-M. Characterization of Electronic and Ionic Transport in
  Li1-{xNi}0.33Mn0.33Co0.33O2({NMC}333) and
  Li1-{xNi}0.50Mn0.20Co0.30O2({NMC}523) as a Function of Li Content.
  \emph{Journal of The Electrochemical Society} \textbf{2016}, \emph{163},
  A1512--A1517\relax
\mciteBstWouldAddEndPuncttrue
\mciteSetBstMidEndSepPunct{\mcitedefaultmidpunct}
{\mcitedefaultendpunct}{\mcitedefaultseppunct}\relax
\EndOfBibitem
\bibitem[Jayawardana \latin{et~al.}(2018)Jayawardana, Carr, Zhao, and
  Majzoub]{Jayawardana2018}
Jayawardana,~W.; Carr,~C.~L.; Zhao,~D.; Majzoub,~E.~H. Voltage-Relaxation
  {GITT} and Reverse Monte Carlo to Determine Lithium Diffusion and
  Distribution in {TiO}2and Highly-Ordered Nanoporous Hard Carbons.
  \emph{Journal of The Electrochemical Society} \textbf{2018}, \emph{165},
  A2824--A2832\relax
\mciteBstWouldAddEndPuncttrue
\mciteSetBstMidEndSepPunct{\mcitedefaultmidpunct}
{\mcitedefaultendpunct}{\mcitedefaultseppunct}\relax
\EndOfBibitem
\bibitem[Shen \latin{et~al.}(2013)Shen, Cao, Rahn, and Wang]{Shen2013}
Shen,~Z.; Cao,~L.; Rahn,~C.~D.; Wang,~C.-Y. Least Squares Galvanostatic
  Intermittent Titration Technique ({LS}-{GITT}) for Accurate Solid Phase
  Diffusivity Measurement. \emph{Journal of The Electrochemical Society}
  \textbf{2013}, \emph{160}, A1842--A1846\relax
\mciteBstWouldAddEndPuncttrue
\mciteSetBstMidEndSepPunct{\mcitedefaultmidpunct}
{\mcitedefaultendpunct}{\mcitedefaultseppunct}\relax
\EndOfBibitem
\bibitem[Hess \latin{et~al.}(2015)Hess, Roode-Gutzmer, Heubner, Schneider,
  Michaelis, Bobeth, and Cuniberti]{Hess2015}
Hess,~A.; Roode-Gutzmer,~Q.; Heubner,~C.; Schneider,~M.; Michaelis,~A.;
  Bobeth,~M.; Cuniberti,~G. Determination of state of charge-dependent
  asymmetric Butler{\textendash}Volmer kinetics for {LixCoO}2 electrode using
  {GITT} measurements. \emph{Journal of Power Sources} \textbf{2015},
  \emph{299}, 156--161\relax
\mciteBstWouldAddEndPuncttrue
\mciteSetBstMidEndSepPunct{\mcitedefaultmidpunct}
{\mcitedefaultendpunct}{\mcitedefaultseppunct}\relax
\EndOfBibitem
\bibitem[Deiss(2005)]{Deiss2005}
Deiss,~E. Spurious chemical diffusion coefficients of Li+ in electrode
  materials evaluated with {GITT}. \emph{Electrochimica Acta} \textbf{2005},
  \emph{50}, 2927--2932\relax
\mciteBstWouldAddEndPuncttrue
\mciteSetBstMidEndSepPunct{\mcitedefaultmidpunct}
{\mcitedefaultendpunct}{\mcitedefaultseppunct}\relax
\EndOfBibitem
\bibitem[Mendoza \latin{et~al.}(2016)Mendoza, Roberts, Brunini, and
  Grillet]{Mendoza2016}
Mendoza,~H.; Roberts,~S.~A.; Brunini,~V.~E.; Grillet,~A.~M. Mechanical and
  Electrochemical Response of a {LiCoO}2 Cathode using Reconstructed
  Microstructures. \emph{Electrochimica Acta} \textbf{2016}, \emph{190},
  1--15\relax
\mciteBstWouldAddEndPuncttrue
\mciteSetBstMidEndSepPunct{\mcitedefaultmidpunct}
{\mcitedefaultendpunct}{\mcitedefaultseppunct}\relax
\EndOfBibitem
\bibitem[Balluffi \latin{et~al.}(2005)Balluffi, Allen, and
  Carter]{Balluffi2005}
Balluffi,~R.~W.; Allen,~S.~M.; Carter,~W.~C. \emph{Kinetics of materials}; John
  Wiley \& Sons, 2005\relax
\mciteBstWouldAddEndPuncttrue
\mciteSetBstMidEndSepPunct{\mcitedefaultmidpunct}
{\mcitedefaultendpunct}{\mcitedefaultseppunct}\relax
\EndOfBibitem
\bibitem[Rickard and Luther(2007)Rickard, and Luther]{Rickard2007}
Rickard,~D.; Luther,~G.~W. Chemistry of Iron Sulfides. \emph{Chemical Reviews}
  \textbf{2007}, \emph{107}, 514--562\relax
\mciteBstWouldAddEndPuncttrue
\mciteSetBstMidEndSepPunct{\mcitedefaultmidpunct}
{\mcitedefaultendpunct}{\mcitedefaultseppunct}\relax
\EndOfBibitem
\bibitem[Fong \latin{et~al.}(1989)Fong, Dahn, and Jones]{Fong1989}
Fong,~R.; Dahn,~J.~R.; Jones,~C. H.~W. Electrochemistry of Pyrite-Based
  Cathodes for Ambient Temperature Lithium Batteries. \emph{Journal of The
  Electrochemical Society} \textbf{1989}, \emph{136}, 3206--3210\relax
\mciteBstWouldAddEndPuncttrue
\mciteSetBstMidEndSepPunct{\mcitedefaultmidpunct}
{\mcitedefaultendpunct}{\mcitedefaultseppunct}\relax
\EndOfBibitem
\bibitem[Evans \latin{et~al.}(2014)Evans, Piper, Kim, Han, Bhat, Oh, and
  Lee]{Evans2014}
Evans,~T.; Piper,~D.~M.; Kim,~S.~C.; Han,~S.~S.; Bhat,~V.; Oh,~K.~H.;
  Lee,~S.-H. Ionic Liquid Enabled {FeS}2for High-Energy-Density Lithium-Ion
  Batteries. \emph{Advanced Materials} \textbf{2014}, \emph{26},
  7386--7392\relax
\mciteBstWouldAddEndPuncttrue
\mciteSetBstMidEndSepPunct{\mcitedefaultmidpunct}
{\mcitedefaultendpunct}{\mcitedefaultseppunct}\relax
\EndOfBibitem
\bibitem[Zou \latin{et~al.}(2020)Zou, Zhao, Wang, Chen, Chen, Ran, Li, Wang,
  Yao, Li, Huang, Niu, and Wang]{Zou2020}
Zou,~J.; Zhao,~J.; Wang,~B.; Chen,~S.; Chen,~P.; Ran,~Q.; Li,~L.; Wang,~X.;
  Yao,~J.; Li,~H.; Huang,~J.; Niu,~X.; Wang,~L. Unraveling the Reaction
  Mechanism of {FeS}2 as a Li-Ion Battery Cathode. \emph{{ACS} Applied
  Materials {\&} Interfaces} \textbf{2020}, \emph{12}, 44850--44857\relax
\mciteBstWouldAddEndPuncttrue
\mciteSetBstMidEndSepPunct{\mcitedefaultmidpunct}
{\mcitedefaultendpunct}{\mcitedefaultseppunct}\relax
\EndOfBibitem
\bibitem[Yersak \latin{et~al.}(2012)Yersak, Macpherson, Kim, Le, Kang, Son,
  Kim, Trevey, Oh, Stoldt, and Lee]{Yersak2012}
Yersak,~T.~A.; Macpherson,~H.~A.; Kim,~S.~C.; Le,~V.-D.; Kang,~C.~S.;
  Son,~S.-B.; Kim,~Y.-H.; Trevey,~J.~E.; Oh,~K.~H.; Stoldt,~C.; Lee,~S.-H.
  Solid State Enabled Reversible Four Electron Storage. \emph{Advanced Energy
  Materials} \textbf{2012}, \emph{3}, 120--127\relax
\mciteBstWouldAddEndPuncttrue
\mciteSetBstMidEndSepPunct{\mcitedefaultmidpunct}
{\mcitedefaultendpunct}{\mcitedefaultseppunct}\relax
\EndOfBibitem
\bibitem[Zhang(2015)]{Zhang2015}
Zhang,~S.~S. The redox mechanism of {FeS}2 in non-aqueous electrolytes for
  lithium and sodium batteries. \emph{Journal of Materials Chemistry A}
  \textbf{2015}, \emph{3}, 7689--7694\relax
\mciteBstWouldAddEndPuncttrue
\mciteSetBstMidEndSepPunct{\mcitedefaultmidpunct}
{\mcitedefaultendpunct}{\mcitedefaultseppunct}\relax
\EndOfBibitem
\bibitem[Son \latin{et~al.}(2013)Son, Yersak, Piper, Kim, Kang, Cho, Suh, Kim,
  Oh, and Lee]{Son2013}
Son,~S.-B.; Yersak,~T.~A.; Piper,~D.~M.; Kim,~S.~C.; Kang,~C.~S.; Cho,~J.~S.;
  Suh,~S.-S.; Kim,~Y.-U.; Oh,~K.~H.; Lee,~S.-H. A Stabilized
  {PAN}-{FeS}2Cathode with an {EC}/{DEC} Liquid Electrolyte. \emph{Advanced
  Energy Materials} \textbf{2013}, \emph{4}, 1300961\relax
\mciteBstWouldAddEndPuncttrue
\mciteSetBstMidEndSepPunct{\mcitedefaultmidpunct}
{\mcitedefaultendpunct}{\mcitedefaultseppunct}\relax
\EndOfBibitem
\bibitem[Amin \latin{et~al.}(2015)Amin, Ravnsb{\ae}k, and Chiang]{Amin2015}
Amin,~R.; Ravnsb{\ae}k,~D.~B.; Chiang,~Y.-M. Characterization of Electronic and
  Ionic Transport in Li1-{xNi}0.8Co0.15Al0.05O2({NCA}). \emph{Journal of The
  Electrochemical Society} \textbf{2015}, \emph{162}, A1163--A1169\relax
\mciteBstWouldAddEndPuncttrue
\mciteSetBstMidEndSepPunct{\mcitedefaultmidpunct}
{\mcitedefaultendpunct}{\mcitedefaultseppunct}\relax
\EndOfBibitem
\bibitem[Noh \latin{et~al.}(2013)Noh, Youn, Yoon, and Sun]{Noh2013}
Noh,~H.-J.; Youn,~S.; Yoon,~C.~S.; Sun,~Y.-K. Comparison of the structural and
  electrochemical properties of layered Li[{NixCoyMnz}]O2 (x~=~1/3, 0.5, 0.6,
  0.7, 0.8 and 0.85) cathode material for lithium-ion batteries. \emph{Journal
  of Power Sources} \textbf{2013}, \emph{233}, 121--130\relax
\mciteBstWouldAddEndPuncttrue
\mciteSetBstMidEndSepPunct{\mcitedefaultmidpunct}
{\mcitedefaultendpunct}{\mcitedefaultseppunct}\relax
\EndOfBibitem
\bibitem[Chouchane \latin{et~al.}(2020)Chouchane, Primo, and
  Franco]{Chouchane2020}
Chouchane,~M.; Primo,~E.~N.; Franco,~A.~A. Mesoscale Effects in the Extraction
  of the Solid-State Lithium Diffusion Coefficient Values of Battery Active
  Materials: Physical Insights from 3D Modeling. \emph{The Journal of Physical
  Chemistry Letters} \textbf{2020}, \emph{11}, 2775--2780\relax
\mciteBstWouldAddEndPuncttrue
\mciteSetBstMidEndSepPunct{\mcitedefaultmidpunct}
{\mcitedefaultendpunct}{\mcitedefaultseppunct}\relax
\EndOfBibitem
\bibitem[LEMEHAUTE \latin{et~al.}(1981)LEMEHAUTE, BREC, DUGAST, and
  ROUXEL]{LEMEHAUTE1981}
LEMEHAUTE,~A.; BREC,~R.; DUGAST,~A.; ROUXEL,~J. The {LixFeS}2 electrochemical
  system. \emph{Solid State Ionics} \textbf{1981}, \emph{3-4}, 185--189\relax
\mciteBstWouldAddEndPuncttrue
\mciteSetBstMidEndSepPunct{\mcitedefaultmidpunct}
{\mcitedefaultendpunct}{\mcitedefaultseppunct}\relax
\EndOfBibitem
\end{mcitethebibliography}

\end{document}